\documentclass[12pt,epsf]{article}
\usepackage{graphicx}
\usepackage{amssymb}
\usepackage{amsmath}
\begin{document}
%%%%%%%%%%%%%%%%%%%%%%%%%%%%%%%%%%%%%%%%%%%

\def\a{\alpha}
\def\b{\beta}
\def\c{\varepsilon}
\def\d{\delta}
\def\e{\epsilon}
\def\f{\phi}
\def\g{\gamma}
\def\h{\theta}
\def\k{\kappa}
\def\l{\lambda}
\def\m{\mu}
\def\n{\nu}
\def\p{\psi}
\def\q{\partial}
\def\r{\rho}
\def\s{\sigma}
\def\t{\tau}
\def\u{\upsilon}
\def\v{\varphi}
\def\w{\omega}
\def\x{\xi}
\def\y{\eta}
\def\z{\zeta}
\def\D{\Delta}
\def\G{\Gamma}
\def\H{\Theta}
\def\L{\Lambda}
\def\F{\Phi}
\def\P{\Psi}
\def\S{\Sigma}

\def\o{\over}
\def\beq{\begin{eqnarray}}
\def\eeq{\end{eqnarray}}
\newcommand{\gsim}{ \mathop{}_{\textstyle \sim}^{\textstyle >} }
\newcommand{\lsim}{ \mathop{}_{\textstyle \sim}^{\textstyle <} }
\newcommand{\vev}[1]{ \left\langle {#1} \right\rangle }
\newcommand{\bra}[1]{ \langle {#1} | }
\newcommand{\ket}[1]{ | {#1} \rangle }
\newcommand{\EV}{ {\rm eV} }
\newcommand{\KEV}{ {\rm keV} }
\newcommand{\MEV}{ {\rm MeV} }
\newcommand{\GEV}{ {\rm GeV} }
\newcommand{\TEV}{ {\rm TeV} }
\def\diag{\mathop{\rm diag}\nolimits}
\def\Spin{\mathop{\rm Spin}}
\def\SO{\mathop{\rm SO}}
\def\O{\mathop{\rm O}}
\def\SU{\mathop{\rm SU}}
\def\U{\mathop{\rm U}}
\def\Sp{\mathop{\rm Sp}}
\def\SL{\mathop{\rm SL}}
\def\tr{\mathop{\rm tr}}

\def\IJMP{Int.~J.~Mod.~Phys. }
\def\MPL{Mod.~Phys.~Lett. }
\def\NP{Nucl.~Phys. }
\def\PL{Phys.~Lett. }
\def\PR{Phys.~Rev. }
\def\PRL{Phys.~Rev.~Lett. }
\def\PTP{Prog.~Theor.~Phys. }
\def\ZP{Z.~Phys. }

\newcommand{\ds}{\displaystyle}

%%%%%%%%%%%%%%%%%%%%%%%%%%%%%%%%%%%%%%%%%%%%%%%%%%%%%%%%%%%%%%%%%%%%

\baselineskip 0.7cm

\begin{titlepage}

\begin{flushright}
KEK-TH-1418 \\
IPMU-10-0189  \\
\end{flushright}

\vskip 1.35cm
\begin{center}
{\large \bf
 A theory of extra radiation in the Universe
}
\vskip 1.2cm
Kazunori Nakayama$^a$,
Fuminobu Takahashi$^b$
and 
Tsutomu T. Yanagida$^{b,c}$

\vskip 0.4cm

{ \it $^a$Theory Center, KEK, 1-1 Oho, Tsukuba, Ibaraki 305-0801, Japan}\\
{\it $^b$Institute for the Physics and Mathematics of the Universe,
University of Tokyo, Kashiwa 277-8568, Japan}\\
{\it $^c$Department of Physics, University of Tokyo, Tokyo 113-0033, Japan}

\vskip 1.5cm

\abstract{
Recent cosmological observations,  such as the measurement of the primordial $^4$He abundance,
CMB, and large scale structure, give preference to the existence of extra radiation component,
$\Delta N_\nu > 0$. 
The extra radiation may be accounted for by particles which were in thermal equilibrium and decoupled
before the big bang nucleosynthesis. Broadly speaking, there are two possibilities: 1) there are
about $10$ particles which have very weak couplings to the standard model particles and 
decoupled much before the QCD phase transition; 2) there is one or a few light particles with a 
reasonably strong coupling to the plasma and it decouples after the QCD phase transition.
Focusing on the latter case, we find that a light chiral fermion is a suitable candidate, which evades 
astrophysical constraints. Interestingly, such a scenario may be confirmed at the LHC.
As a concrete example, we show that such a light fermion naturally appears in the $E_6$-inspired GUT.
}
\end{center}
\end{titlepage}

\setcounter{page}{2}

%%%%%%%%%%%%%%%%%%%%%%%%%%%%%%%%%%%%%%%%%%%%%
\section{Introduction}
%%%%%%%%%%%%%%%%%%%%%%%%%%%%%%%%%%%%%%%%%%%%%

One of the most important discoveries in cosmology is that the Universe is expanding. 
According to general relativity, the expansion rate is determined by the energy contained in the Universe.
In fact, measuring the expansion rate has been one of the central issues in cosmology, because it provides fundamental cosmic
age and distance scales. For instance, a precise measurement of  Type Ia supernovae (SNe) revealed that the present Universe
is filled with dark energy, which accelerates the cosmic expansion~\cite{Riess:1998cb,Perlmutter:1998np}. 
It is even possible to infer the particle content of the Universe in the past, by measuring the primordial abundance
of $^4$He, the cosmic microwave background (CMB) anisotropies, and large scale structure (LSS).

The primordial abundance of $^4$He was the key observational evidence for the big bang theory.
The big bang nucleosynthesis (BBN) calculation agreed reasonably well with the observed the $^4$He mass fraction $Y_p$
together with other light element abundances given in terms of a function of the baryon-to-photon ratio, $\eta$. In the post-WMAP era, 
$\eta$ was determined to a very high accuracy~\cite{Komatsu:2010fb}, which allowed the internal consistency check
of the BBN calculation based on the standard big bang cosmology.  

It is known that the $Y_p$
is sensitive to the expansion rate of the Universe during the BBN epoch\footnote{
$Y_p$ is also sensitive to large lepton asymmetry, especially of the electron type, if any~\cite{Enqvist:1990ad,Foot:1995qk,Shi:1996ic,MarchRussell:1999ig,Kawasaki:2002hq,Ichikawa:2004pb}.
}, while it is a rather insensitive baryometer. In the post-WMAP era, it turned out that the helium mass fraction determined by extragalactic HII regions was 
smaller than the value of $Y_p$ predicted by the BBN calculation using the WMAP determined $\eta$. The apparent tension was partly due to the underestimated
error associated with the early determinations of $Y_p$, and it was pointed out that
the precise determination of the helium abundance is limited by systematic uncertainties~\cite{Olive:2004kq}. 
Since then, as the physical processes as well as the associated systematic corrections have been studied thoroughly,
the estimated helium mass fraction has increased substantially. Recently, the authors of Ref.~\cite{Izotov:2010ca} claimed an excess of $Y_p$ at the  $2 \sigma$ level,
$Y_p = 0.2565 \pm 0.0010\, ({\rm stat}) \pm 0.0050\, ({\rm syst})$, which can be 
understood in terms of an effective number of light neutrino species as $N_{\rm eff} = 3.68^{+0.80}_{-0.70}$ $(2 \sigma)$. 
For comparison, the WMAP value is given by $Y_p = 0.2486 \pm 0.0002$\,$(68\%{\rm CL})$~\cite{Cyburt:2008kw},
and the standard cosmology gives $N_{\rm eff} = 3.046$. 
On the other hand, the authors of Ref.~\cite{Aver:2010wq} estimated the primordial helium abundance with an unrestricted Monte Carlo taking account of all systematic corrections and obtained
$Y_p = 0.2561 \pm 0.0108$\,$(68\%{\rm CL})$, which is in broad agreement with the WMAP result. Although the precise determination of the helium abundance is still limited 
by systematic uncertainties, it is intriguing that the center values of the two different results agree with each other and both are higher than the WMAP value.

The combined analysis of the CMB and LSS data also constrains the relativistic species 
after the matter-radiation equality. The analysis based on WMAP 7yr +BAO+H$_0$ gives
$N_{\rm eff} = 4.34^{+0.86}_{-0.88}$\,$(68\%{\rm CL})$~\cite{Komatsu:2010fb}.
By combining the latest result of the Atacama Cosmology Telescope (ACT), the constraint is slightly improved and becomes
$N_{\rm eff} = 4.56 \pm 0.75$\,$(68\%{\rm CL})$~\cite{Dunkley:2010ge}. Thus, the CMB and LSS data suggests the presence
of extra radiation at the $2\sigma$ level. 

It is remarkable that, while the helium abundance, the CMB and LSS data are sensitive
to the expansion rate of the Universe at vastly different times, all the data mildly favor
additional relativistic species, $\Delta N_{\rm eff} \sim 1$. Although it may be still premature to draw a definite conclusion,
we assume in the following, that there is indeed extra radiation
suggested by the current observations, and study its implications for particle physics.

The extra radiation may be ``dark" radiation composed of unknown light and relativistic particles, $X_i$, where the subscript $i = 1 \cdots n$
labels different species. Among various possibilities of production mechanisms, 
we focus on a scenario that $X_i$ were in thermal equilibrium in the early Universe. This is because
some amount of fine-tuning is necessary to account for $\Delta N_{\rm eff} \sim 1$ in such a case that
 $X_i$ is produced non-thermally by  the decay of heavy particles~\cite{Ichikawa:2007jv}. If we assume that $X_i$
is in equilibrium, their total abundance is determined by the number of species $n$ and the decoupling time.\footnote{
If $X_i$ has a renormalizable coupling with the SM particles, $X_i$ enters thermal equilibrium at a late time.
We will discuss this case in Sec.~\ref{sec:2-1}.
}
In particular,  the abundance of $X_i$ that has decoupled before the QCD phase transition is diluted by a factor
of $5 \sim 10$. Therefore, broadly speaking,  there are two possibilities: (1)
there are $5 \sim 10$ light particles in equilibrium and decoupled  before the QCD phase transition, or (2) 
there is one or  a few light particles in equilibrium which has decoupled after the QCD transition before the BBN epoch. 
In the former case, $X_i$ has only suppressed couplings to the standard model (SM) sector, and it is in general difficult
to study the properties of dark radiation by experiments. Also a mild fine-tuning is necessary to account for $\Delta N_{\rm eff} \sim 1$.\footnote{
\label{ft}
For instance, there might be a hidden SU(N) gauge symmetry and a massive matter field in the bifundamental representation of the SU(N) and
SM gauge groups. Assuming that the mass is heavier than the weak scale, the SU(N) gauge bosons can account for  $\Delta N_{\rm eff} \sim 1$
if the SU(N) gauge coupling is sufficiently small and if $N=3 \sim 4$. If the mass of the matter happens to be at the weak scale, it may be within the reach
of the LHC.
}
We focus on the latter case because it can naturally explains the deviation from the standard value, $\Delta N_{\rm eff} \sim 1$,
and because, as we shall see, it leads to interesting implications for the LHC. For simplicity we assume $n=1$ unless otherwise stated.

In order to account for the excess $\Delta N_{\rm eff} \sim 1$ suggested by the CMB and LSS data, the mass of $X_1$
must be  lighter than $0.1$\,eV.\footnote{Note that the temperature of $X_1$ is in general lower than the CMB photon temperature
by at most $\sim 4$. The bound on the mass is still valid, taking account of this factor. FT thanks J. Redondo for pointing out this issue.
} Such a light mass is a puzzle and clearly calls for some explanation.
The light mass could be a result of an underlying symmetry such as gauge symmetry, shift symmetry, or  chiral
symmetry. In the following section we consider each case and discuss possible astrophysical constraints. As we shall see,
a chiral fermion, coupled to the standard model with interactions suppressed by the TeV scale, is a viable candidate
for dark radiation. Interestingly enough, such a scenario may be within the reach of the LHC.

%%%%%%%%%%%%%%%%%%%%%%%%%%%%%%%%%%%%%%%%%%%%%
\section{Candidates for the extra radiation}
%%%%%%%%%%%%%%%%%%%%%%%%%%%%%%%%%%%%%%%%%%%%%

In this section we discuss various possibilities for generating the extra radiation, 
$\Delta N_{\rm eff} \simeq 1$, carried by a light particle $X$.
In order for  $X$  to be regarded as radiation both at the BBN and CMB epochs,
its mass must be smaller than $\sim 0.1$~eV. One explanation for such a small mass is that
there is a symmetry forbidding the mass.\footnote{
We do not consider sterile (right-handed) neutrinos~\cite{Hamann:2010bk} here because there is no symmetry to keep their mass 
light, especially in the context of the seesaw mechanism and the leptogenesis scenario.
}  We consider the following three symmetries: gauge symmetry,
shift symmetry, and chiral symmetry.
First,  a gauge symmetry forbids a bare mass of the gauge boson. While
a scalar mass is not protected by symmetries in general,\footnote{
	Supersymmetry (SUSY) helps to obtain a light scalar boson, since it relates a
	scalar boson to its fermionic superpartner. However SUSY must be spontaneously broken,
	and this generically induces a scalar boson mass heavier than $0.1$\,eV.
}
there is an important exception, that is, a Nambu-Goldstone (NG) boson, which appears in association with the spontaneous breakdown of a global symmetry.
If the  global symmetry is exact, the NG boson remains massless because of the shift symmetry.
Finally, a chiral symmetry forbids the bare mass term of a chiral fermion.
In the following we consider these possibilities separately.

The energy density of the extra radiation consisting of a particle $X$ is related to
 the effective number of neutrino species as
\begin{equation}
	\Delta N_{\rm eff} = \frac{\rho_{X}}{\rho_\nu} 
	=\epsilon \left( \frac{g_{*\nu}}{g_{*X}} \right)^{4/3},
\end{equation}
where
\begin{equation}
	\epsilon =\left \{ \begin{array}{ll}
	4/7 & ~~~{\rm for~a~NG~boson} \\
	1 & ~~~{\rm for~a~chiral~fermion} \\
	8/7    & ~~~{\rm for~a~massless~U(1)~gauge~boson} 
	\end{array}
	\right. .
\end{equation}
Here $\rho_\nu (\rho_{X})$ and $g_{*\nu} (g_{*X})$ are the energy density of one neutrino species($X$)
and relativistic degrees of freedom $g_*$ evaluated at the time of decoupling of neutrinos ($X$) from thermal plasma.
The relativistic degrees of freedom drops sharply from about $50$ to $20$ at $T \sim 200$\,MeV during the QCD phase transition.
In the standard cosmology, the neutrinos decouple at a temperature about a few MeV, and
 $g_{*\nu}$ is equal to $10.75$. 
The helium abundance is sensitive the expansion rate at the decoupling of neutrinos. 
After the neutrino decoupling, $g_*$ becomes $3.36$ for $T \ll $\,MeV.
Thus, if the decoupling temperature of $X$ is between a few MeV and $100$\,MeV,
$g_{*X}$ is comparable to $g_{*\nu}$, and one can naturally explain the extra radiation $\Delta N_{\rm eff} \sim 1$.
In the following we focus on this case, since
otherwise $\Delta N_{\rm eff}$ becomes either smaller or bigger than $1$, which would necessitate multiple species of $X$ or
non-thermal production to achieve $\Delta N_{\rm eff} \sim 1$.

\subsection{Spin 1 - Gauge bosons}
\label{sec:2-1}

We consider a U(1) gauge symmetry (see footnote \ref{ft} for the case of non-Abelian gauge group).
The general Lagrangian for a hidden photon $\gamma'$ with a mass of $m_{\gamma'}$ is given by
\begin{equation}
	\mathcal L = -\frac{1}{4}F_{\mu\nu}F^{\mu\nu}-\frac{1}{4}B_{\mu\nu}B^{\mu\nu}
	+\frac{\chi}{2}F_{\mu\nu}B^{\mu\nu}+\frac{1}{2}m_{\gamma'}^2 B_\mu B^\mu,
\end{equation}
where $B_\mu$ and $B_{\mu\nu}$ denote the hidden photon field and its field strength.
The third term represents a kinetic mixing between $\gamma$ and $\gamma'$,
and $\chi$ is a numerical coefficient. In this example, since the hidden photon has a renormalizable coupling
to the usual photon, it enters thermal equilibrium at a late time, which should be contrasted to the other two cases
considered in the following, where the abundance of extra radiation is fixed after the decoupling.

By redefining $A_\mu$ and $B_\mu$ as $A_\mu' = (1-\chi^2)^{1/2} A_\mu$ and $B_\mu' = B_\mu - \chi A_\mu,$
we can remove the kinetic mixing term and obtain canonical kinetic terms.
Then, there appears a photon-hidden photon mixing in the mass matrix as
\begin{equation}
	\mathcal M^2 = m_{\gamma'}^2\begin{pmatrix}
		\chi^2/(1-\chi^2)          && \chi / \sqrt{1-\chi^2} \\
		\chi  / \sqrt{1-\chi^2}    &&  1
	\end{pmatrix},
\end{equation}
where $\chi$ satisfies $|\chi| < 1$.
Note that det$\mathcal M^2=0$ and hence there is a massless eigenstate, which can be
regarded as a photon.
In this setup we can calculate the probability for a photon to be converted into a hidden photon in  
thermal plasma.
If the effective rate $\Gamma$ for producing a hidden photon by the process like $\gamma e \to \gamma' e$
exceeds the Hubble parameter $H$, hidden photons are considered to be thermalized.\footnote{
	In the situation $m_{\gamma'} \sim m_\gamma$, there is a resonant enhancement 
	of the conversion rate, which is analogous to the MSW resonance in neutrino oscillations.
}
Here the rate $\Gamma$ is estimated as~\cite{Jaeckel:2008fi,Jaeckel:2010ni},
\begin{equation}
	\Gamma \;\simeq \;
	\left \{ \begin{array}{ll}
		{\ds \chi^2 \left( \frac{m_{\gamma'}}{m_{\gamma}} \right)^4 \Gamma_{\rm C} }
			& {\rm ~~for~}m_{\gamma'} \ll m_\gamma   \\
			&\\
		\chi^2 \Gamma_{\rm C} 
			& {\rm ~~for~}m_{\gamma'} \gg m_\gamma  
	\end{array}
	\right.,
\end{equation}
where $m_\gamma$ is the effective photon mass in the medium, and $\Gamma_{\rm C}\sim \alpha_e^2 T$ is the
Compton scattering rate, and $\chi \ll 1$ is assumed. 
In the early Universe with $T\gtrsim 1\,$MeV, a photon has a thermal mass $m_\gamma \sim T$,
while the hidden photon mass must be smaller than $\sim 0.1$~eV
for it to be regarded as the extra radiation. One can easily check that 
 the hidden photons are never thermalized before the BBN epoch
for $m_{\gamma'} \lesssim 0.1$eV for $|\chi | < 1$. Thus, this scenario cannot explain
$\Delta N_{\rm eff} \sim 1$ at the BBN epoch.\footnote{
It is possible to obtain $\Delta N_{\rm eff} \sim 1$ 
	at the CMB epoch for $\chi \sim 10^{-5}$, using the resonant enhancement of the conversion 
	for a certain choice of $m_{\gamma'}\sim 1$meV~\cite{Jaeckel:2008fi}.
	However, one cannot explain $\Delta N_{\rm eff}\sim 1$ {\it both} at the BBN and CMB epochs.
	Also, the helium abundance is decreased because of the enhancement of $\eta$ after the BBN,
	which will make the agreement with the observation worse.
}

\subsection{Spin 0 - NG bosons}

Next, let us consider a NG boson $a$, which is associated with the spontaneous breakdown of 
some global symmetry at a scale of $f_a$.
We assume that the explicit breaking of the symmetry is so small that the $a$ 
remains practically massless. The value of $N_{\rm eff}$ is at most $4/7$ in this scenario.

First, consider an axion-like particle coupled to photons as 
\begin{equation}
	\mathcal L = \frac{\alpha_e}{8\pi}\frac{a}{f_a}F_{\mu\nu}\tilde F^{\mu\nu}.
\end{equation}
The freeze-out temperature is determined by the balance between the Hubble expansion rate,
$H\sim T^2/ M_P$, and the rate of processes such as $\gamma e \leftrightarrow ae$,
where $M_P \simeq 2.4\times 10^{18}$\,GeV is the reduced Planck mass.
The latter rate is evaluated as
\begin{equation}
	\Gamma (\gamma e \leftrightarrow ae) \sim \langle \sigma v\rangle n_e
	\sim \frac{\alpha_e^3 T^3}{f_a^2},
\end{equation}
for $T\gtrsim 1$~MeV,
where $\langle \sigma v\rangle $ is the cross section of the corresponding process
and $n_e$ denotes the electron number density.
We then find that the freeze-out temperature of $a$ is $T_f \sim 10{\rm MeV}(f_a/10^5{\rm GeV})^2$.
Thus we need $f_a \sim 10^5$GeV to account for the extra radiation. However, it is
excluded by the constraints from the cooling of stars and white dwarfs. 
For example, the constraint from the observation of horizontal branch (HB) stars  gives
$f_a \gtrsim 10^8$GeV~\cite{Raffelt}. 

Next let us consider a case that $a$ interacts with hadrons, as in the case of QCD axion
~\cite{Kim:1986ax}.
The interaction Lagrangian is given by 
\begin{equation}
	\mathcal L = \frac{\alpha_e C_{a\gamma\gamma}}{8\pi}\frac{a}{f_a}F_{\mu\nu}\tilde F^{\mu\nu}
	+ \frac{\alpha_s}{8\pi}\frac{a}{f_a}F_{\mu\nu}^a\tilde F^{\mu\nu a}
	 + \frac{a}{f_a}im_q \bar q \gamma_5 q.
\end{equation}
It is possible that the axion-photon-photon coupling $C_{a\gamma\gamma}$
is small due to an accidental cancellation in the hadronic axion model~\cite{Chang:1993gm,Moroi:1998qs}.
In this case constraints from HB stars can be avoided,
while the freeze-out temperature is comparable to or higher than $\mathcal O(10)$~MeV due to the 
axion-hadron coupling for $f_a \lesssim 10^8$~GeV~\cite{Masso:2002np,Hannestad:2005df}.
On the other hand, the observation of SN1987A constrains $f_a$ as
$10^5$~GeV $\lesssim f_a \lesssim 10^6$~GeV due to the axion-hadron 
interaction term~\cite{Moroi:1998qs}, 
which translates into the axion mass of $1-10$~eV, the so-called ``hadronic axion window''.
Therefore the axion mass is too heavy for the axion to be regarded as an extra radiation at the CMB epoch;
such axions should be rather regarded as a hot dark matter component.
Including the hot dark matter component does not improve a fit to observational data.
In fact, it was recently shown in Ref.~\cite{Hannestad:2010yi}  that the axion mass is constrained as
$m_a < 1\,$eV by the CMB and LSS data. Thus it is difficult to explain $\Delta N_{\rm eff} \sim 1$ in this scenario.

\subsection{Spin 1/2 - Chiral fermions}  \label{sec:chiral}

Let us consider a chiral fermion $\psi$, which is assumed to have a non-vanishing charge of a new U(1) gauge symmetry
and hence its mass term is forbidden.\footnote{
	Following discussions do not much depend on whether a new gauge boson
	is associated with an Abelian or non-Abelian gauge group.
	Here we consider the case of U(1) gauge boson for simplicity.
} It is coupled to the gauge boson $A_H^\mu$, as
$\mathcal L_{\rm int} = ig_{A\psi\psi}A_H^\mu \bar \psi \gamma_\mu \psi$.
We also assume that SM fermions, which we collectively denote by $f$, have interactions
with $A_H^\mu$, as $\mathcal L_{\rm int} = ig_{Aff}A_H^\mu \bar f \gamma_\mu f$. We assume that
the U(1) gauge symmetry is spontaneously broken and therefore the $A_H$ acquires a heavy mass, $m_A$.
Integrating out the heavy gauge boson,
we obtain an effective  four fermion interaction as
\begin{equation}
	\mathcal L _{\rm eff} = \frac{1}{\Lambda^2}(\bar f \gamma^\mu f )(\bar \psi \gamma_\mu \psi),
	\label{4f}
\end{equation}
where $\Lambda^2 = m_A^2 g_{A\psi\psi}^{-1}g_{Aff}^{-1}$ is the cutoff scale.

The freeze-out temperature of $\psi$ is determined again by the balance between the Hubble expansion rate,
$H\sim T^2/ M_P$, and the rate of the interaction such as $e^+e^- \leftrightarrow \psi\psi$.
The latter is evaluated as
\begin{equation}
	\Gamma (e^+e^- \leftrightarrow \psi\psi) \sim \langle \sigma v\rangle n_e
	\sim \frac{T^5}{\Lambda^4},
\end{equation}
for $T \gtrsim 1$~MeV,
where $\langle \sigma v\rangle $ is the cross section of the corresponding process.
The freeze-out temperature of $\psi$ is then evaluated roughly as
$T_f \sim 10$MeV$(\Lambda/{\rm TeV})^{4/3}$.
Thus the light chiral fermion with a cutoff scale of $\mathcal O(1){\rm TeV}$ is a prime
candidate for the extra radiation.

Let us discuss astrophysical constraints on such $\psi$.
Clearly, if $\Lambda \gtrsim 1$TeV, the coupling of $\psi$ with electrons
is weaker than the usual weak interactions, 
and hence the cooling of stars through the emission of $\psi$ is not efficient.
The energy loss rate of stars due to emissions of $\psi$ is suppressed by the factor $G_F^{-2} / \Lambda^4\sim 0.01\, (\Lambda/{\rm TeV})^{-4}$ 
compared with that by the neutrino emission, where $G_F$ is the Fermi constant.
Therefore constraints from stars can be easily evaded. On the other hand, the  supernova cooling argument 
places a much tighter constraint. This is because the neutrinos are trapped inside
a supernova because of its extremely high density and temperature, and such $\psi$ may carry a significant amount of energy
from the supernova.
 The constraint on $\Lambda$ from the supernova cooling argument reads 
$G_F^{-1}/\Lambda^2 \lesssim 3 \times 10^{-3}$, namely, $\Lambda \gtrsim 6$\,TeV~\cite{Raffelt:1999tx}.
For $\Lambda \sim 6$\,TeV, the freeze-out temperature is about $100$\,MeV, and the relativistic degree of freedom 
is given by $g_{* X} \simeq 20$. Thus, in order to explain $\Delta N_{\rm eff} \sim 1$, a couple of such $\psi$'s are needed.

We may simply assume the presence of such a new U(1), but it
may be a part of a large gauge group of the  grand unified theory (GUT). In the next section we give one example inspired by the $E_6$ GUT, which fulfills the required property. In particular, there are three light chiral fermions in this model.

%%%%%%%%%%%%%%%%%%%%%%%%%%%%%%%%%%%%%%%%%%%%%
\section{An example : $E_6$-inspired GUT}
%%%%%%%%%%%%%%%%%%%%%%%%%%%%%%%%%%%%%%%%%%%%%

We have seen that a chiral fermion is a suitable candidate for the extra radiation of the Universe.
Now we discuss a possible origin of the new U(1) gauge symmetry and the extra fermion.

We need an additional gauge symmetry to forbid a bare mass for a chiral fermion, and
the simplest one is a  U(1) gauge symmetry. The U(1) symmetry 
must be spontaneously broken at TeV scale to produce the right abundance of extra radiation.
An important constraint on such U(1) is that it must be free from the quantum anomaly. One of the anomaly-free U(1)s is
U(1)$_{B-L}$, which naturally appears in the SO(10) GUT.
Actually, however, the U(1)$_{B-L}$ symmetry should be spontaneously broken at  a scale much higher than the weak scale,
in order to explain tiny neutrino masses through the seesaw mechanism~\cite{seesaw}.
Then, we need to enlarge the gauge group, 
and in fact, an additional anomaly-free U(1) often appears in the breaking pattern of a GUT gauge group with a higher rank.

%%%%%%%%%%%%%%%% table %%%%%%%%%%%%%%%%%%%%%%
\begin{table}[t]
  \begin{center}
    \begin{tabular}{ | c | c |}
      \hline 
         $SO(10)\times U(1)_\psi$  & $SU(5)\times U(1)_\psi \times U(1)_\chi$  \\
       \hline 
        ~                                               & $\psi_{\bf 10}^{\rm (SM)}(1,1)$ \\
        $\Psi_{\bf 16}(1)$                  &  $\psi_{\bf \bar 5}^{\rm (SM)}(1,-3)$ \\
        ~					 &  $\psi_{\bf 1}^{\rm (SM)}(1,5)$ $=\nu_{\rm R}$ \\
      \hline
        $\Psi_{\bf 10}(-2)$                & $\psi_{\bf 5}^{\bf (10)}(-2,-2)$ \\
        ~			                   &  $\psi_{\bf \bar 5}^{\bf (10)}(-2,2)$ \\
      \hline 
         $\Psi_{\bf 1}(4)$                   & $\psi_{\bf 1}(4,0)$ \\
       \hline
       ~                                               & $\phi_{\bf 10}^{\bf (16)}(1,1)$ \\
        $\Phi_{\bf 16}(1)$                  &  $\phi_{\bf \bar 5}^{\bf (16)}(1,-3)$ \\
        ~					 &  $\phi_{\bf 1}^{\bf (16)}(1,5)$\\
      \hline
        $\Phi_{\bf 10}(-2)$                & $\phi_{\bf 5}(-2,-2)$    $\supset$ SM Higgs\\
        ~			                   &  $\phi_{\bf \bar 5}(-2,2)$ $\supset$ SM Higgs\\
      \hline 
         $\Phi_{\bf 1}(4)$                   & $\phi_{\bf 1}(4,0)$  $=\phi_X$ \\
        \hline
    \end{tabular}
    \caption{ 
    		Notation and charge assignments on the fields in the model.
           }
    \label{table}
  \end{center}
\end{table}
%%%%%%%%%%%%%%%%%%%%%%%%%%%%%%%%%%%%%%%%%%%%%% 

Here we consider a gauge group of two additional anomaly-free U(1)'s,
$SU(5)\times U(1)_\psi\times U(1)_\chi$, where SU(5) includes the SM gauge groups.
This is inspired by the $E_6$ model of the GUT~\cite{Langacker:2008yv}, 
since it has a symmetry breaking pattern,
$E_6 \to SO(10)\times U(1)_\psi$ and $SO(10)\to SU(5)\times U(1)_\chi$.
Although our result does not depend on the details of underlying higher rank GUT theory,
we use the $E_6$ notation in the following analyses for concreteness. 

The $E_6$ group has a ${\bf 27}$ representation, which can be decomposed as
${\bf 27 = 16_1 + 10_{-2} +1_{4}}$ in terms of the SO(10) representation,
where the subscript denotes the U(1)$_\psi$ charge. Let us take a fermion $\Psi_{\bf 27}$ in a ${\bf 27}$ representation,
which contains $\Psi_{\bf 16}$, $\Psi_{\bf 10}$ and $\Psi_{\bf 1}$.
Then, $\Psi_{\bf 16}$ contains all the SM fermions in one generation
as well as a SM singlet fermion which is identified with a right-handed neutrino.
Note that ${\bf 16}$  is decomposed as ${\bf 16}= {\bf 10}+{\bf \bar 5}+{\bf 1}$ in terms of the SU(5) representation.
See Table~\ref{table} for notation and charge assignments.
We also introduce a scalar $\Phi_{\bf 27}$ in a ${\bf 27}$ representation, which contains $\Phi_{\bf 16}$, $\Phi_{\bf 10}$ and $\Phi_{\bf 1}$.
The $\Phi_{\bf 10}$ is the Higgs field in the ${\bf 10_{\rm H}}$ 
($= {\bf 5}_{\rm H}+{\bf \bar 5}_{\rm H}$ in terms of SU(5) representation), 
which contains the SM Higgs boson.  
All the SM Yukawa couplings arise from $\Phi_{\bf 27} \Psi_{\bf 27} \Psi_{\bf 27}$.
For instance the Yukawa couplings associated with the SM fermions $\Psi_{\bf 16}$
can be written as $\Phi_{\bf 10} \Psi_{\bf 16} \Psi_{\bf 16}$.
On the other hand, the right-handed neutrino, $\psi_1 = \nu_{\rm R}$, 
which is a SU(5) singlet of $\Psi_{\bf 16}$, obtains a mass from an interaction,
$ \Phi_{\overline{\bf 126} }\Psi_{\bf 16}\Psi_{\bf 16}$, where $\Phi_{\overline{\bf 126} }$ is
a SO(10) ${\bf \overline {126}}$ representation Higgs,  a part of the {\bf 351'} representation of $E_6$~\cite{Slansky:1981yr}.
If the singlet part of $\Phi_{\overline{\bf 126}}$ develops a large vacuum expectation value (VEV), 
the right-handed neutrino acquires a Majorana mass of $\sim \langle \Phi_{\overline{\bf 126}} \rangle$.
Note that the VEV leaves $5U(1)_\psi - U(1)_\chi$ unbroken, which we call $U(1)_X$ in the following.

The singlet fermion $\Psi_{\bf 1} = \psi_{\bf 1}$ remains massless as long as $U(1)_X$ is unbroken.
Let us assume that the $U(1)_X$ is spontaneously broken only by the non-vanishing VEV of $\phi_X$, $\langle \phi_{X} \rangle = \xi$,
where $\phi_X$ is the SO(10) singlet Higgs scalar.
The U(1)$_X$ gauge boson then acquires a mass of the order of $\xi$. In order to have
a cutoff scale $\Lambda$ of $\mathcal O(1)$TeV in the four fermion interaction (\ref{4f}), hereafter we will set $\xi = \mathcal O(1)$TeV.
The $\psi_{\bf 1}$ obtains a mass only through the following higher dimensional operator,
\begin{equation}
	\mathcal L \;\sim\;
	%\frac{\Phi_{\bf 1}^*	\Phi_{\bf 1}^* \Psi_{\bf 1} \Psi_{\bf 1}}{M} + {\rm h.c.}.
	\frac{\phi_{X}^*	\phi_{X}^* \psi_{\bf 1} \psi_{\bf 1}}{M} + {\rm h.c.},	
\end{equation}
where the form of the interaction is determined by the U(1)$_\psi$ charge conservation. Thus,
$\psi_{\bf 1}$ obtains a mass $m \sim \xi^2 / M \sim 10^{-3}$\,eV, where we have substituted 
$\xi= 1\,$TeV and $M =  M_P$.
If there is such a singlet fermion $\psi_{\bf 1}$ in each generation, 
we would have $\Delta N_{\rm eff} \sim 1$ under the constraint $\Lambda \gtrsim 6$TeV
discussed in Sec.~\ref{sec:chiral}.
%we would have $\Delta N_{\rm eff} = 3$. It is also possible
%to realize $\Delta N_{\rm eff} \sim 1$ by slightly increasing the scale of $\xi$.
Thus this singlet fermion $\psi_{\bf 1}$  in the $E_6$-inspired GUT is a suitable candidate for the extra radiation.

Let us comment on the fate of  $\Psi_{\bf 10}$, which
acquires a mass $m_{\bf 10_{\rm f}}$ of order of $\xi$
through a coupling to the singlet Higgs,
$\Phi_{\bf 1} \Psi_{\bf 10} \Psi_{\bf 10}$.
The $\Psi_{\bf 10}$ decays into the Higgs and the singlet fermion
through the interaction $\Phi_{\bf 10} \Psi_{\bf 10} \Psi_{\bf 1}$.
Since the SM Higgs has a mass of $\mathcal O (100)$GeV,
the non-colored part of $\Psi_{\bf 10}$ decays quickly into the SM Higgs and $\Psi_{\bf 1}$.
However, the colored Higgs must have a huge mass of order of the GUT scale 
in order to suppress the proton decay, and hence the colored part of the $\Psi_{\bf 10}$
can decay only through the exchange of virtual colored Higgs boson.
The decay rate is suppressed as
\begin{equation}
	\Gamma (\Psi_{\bf 10}^{\rm color}) \sim 10^{-2}\frac{m_{\bf 10_{\rm f}}^5}{M_{\rm GUT}^4}
%	\simeq (7\times 10^{26}~{\rm sec})^{-1}
	\simeq (10^{27}~{\rm sec})^{-1}
	\left( \frac{m_{\bf 10_{\rm f}}}{1{\rm TeV}} \right)^5
	\left( \frac{10^{16}{\rm GeV}}{M_{\rm GUT}} \right)^4.
\end{equation}
Thus its lifetime is much longer than the present age of the Universe, and is regarded as a stable particle.
The relic abundance of colored particles was estimated in Ref.~\cite{Kang:2006yd},
where it was pointed out that they annihilate efficiently after the QCD phase transition
and the abundance is significantly reduced.
However, even such a small amount of stable colored particles may have dangerous effects on
BBN~\cite{Kusakabe:2009jt}.

The cosmological problem of the stable colored particles can be avoided if the reheating temperature of the
Universe after inflation is lower than the mass $m_{\bf 10_{\rm f}} \sim {\rm TeV}$, which however is difficult
to reconcile with the leptogenesis scenario.  Alternatively, we may introduce a small mixing between $\Psi_{\rm 10}$ 
and the SM fermions. Note that there is an interaction,
\beq
{\cal L}\;=\; y \phi_{\bf 1}^{\bf (16)} \psi_{\bf \bar 5}^{{\rm (SM)}}  \psi_{{\bf 5}}^{(\bf 10)}  + {\rm h.c.}, 
\eeq
where $y$ is a coupling constant taken to be real, $\phi_{\bf 1}^{\bf (16)}$ is a part of $\Phi_{\bf 16}$, 
$\psi_{\bf \bar 5}^{{\rm (SM)}}$  and  $\psi_{{\bf 5}}^{(\bf 10)}$ are a part of
$\Psi_{16}$ and $\Psi_{10}$, respectively. If $\phi_{\bf 1}^{\bf (16)} $ develops a tiny VEV of $\langle \phi_{\bf 1}\rangle$, 
the colored part of 
$\Psi_{\bf 10}$ will be mixed with the SM quarks, and therefore it will decay into the SM particles via the mixing.
The decay rate is estimated to be
\begin{equation}
	\Gamma (\Psi_{\bf 10}^{\rm color}) \sim 10^{-2}\theta^2 m_{\bf {10_f}} 
%	\simeq (7\times 10^{-12}~{\rm sec})^{-1} 
	\simeq (10^{-11}~{\rm sec})^{-1} 
	\left( \frac{\theta}{10^{-7}} \right)^2
	\left( \frac{m_{\bf {10_f}} }{1{\rm TeV}} \right),
\end{equation}
where $\theta \simeq y\langle \phi_{\bf 1}\rangle / \xi$ denotes the effective mixing angle.
Note that the VEV of $\phi_{\bf 1}^{\bf (16)}$ breaks $Z_{2(B-L)}$, and 
therefore, it cannot be arbitrarily large, since otherwise
the baryon asymmetry  would be erased before the electroweak phase transition.
Requiring that $\psi_{\bf \bar 5}^{{\rm (SM)}} $-$\psi_{{\bf 5}}^{(\bf 10)}$ conversion process,
whose rate is given by $\Gamma\sim \alpha_s^2 \theta^2 T$, does not reach equilibrium before the electroweak transition,
we find $\theta \lesssim 10^{-7}$~\cite{Campbell:1990fa,Endo:2009cv}. 
The $\Psi_{\bf 10}^{\rm color}$ decays much before the BBN for such a small value of $\theta$.
Interestingly, the decay of $\Psi_{\bf 10}^{\rm color}$ may be observed inside the detector at the LHC.

Another solution is to extend the above framework to a SUSY version.
Then $\Psi_{\rm 10}$ will decay into a SM fermion and SUSY particles
via an exchange of  a fermionic superpartner of $\Phi_{\bf 10}$.
The decay rate of this process is now given by
\begin{equation}
	\Gamma (\Psi_{\bf 10}^{\rm color}) \sim 10^{-2}\frac{m_{\bf 10_{\rm f}}^3}{M_{\rm GUT}^2}
%	\simeq (7~{\rm sec})^{-1}
	\simeq (10~{\rm sec})^{-1}
	\left( \frac{m_{\bf 10_{\rm f}}}{1{\rm TeV}} \right)^3
	\left( \frac{10^{16}{\rm GeV}}{M_{\rm GUT}} \right)^2.
\end{equation}
and hence is much faster than that induced by the dimension six operator,
as long as the $\Psi_{\bf 10}^{\rm color}$ is heavier than the SM particles and their superpartners. 
Thus it can decay before BBN begins, and become cosmologically harmless.
Although this decay process produces SUSY SM particles, their abundance is negligibly small
because the $\Psi_{\bf 10}^{\rm color}$ annihilates efficiently during the QCD phase transition, 
as mentioned above~\cite{Kang:2006yd}.

%%%%%%%%%%%%%%%%%%%%%%%%%%%%%%%%%%%%%%%%%%%%%
\section{Discussion and Conclusions}
%%%%%%%%%%%%%%%%%%%%%%%%%%%%%%%%%%%%%%%%%%%%%

So far we have focused on a case that a light particle $X$ was in equilibrium and decoupled
after the QCD transition. Let us here briefly mention the other possibility, namely, the $X$
decoupled before the QCD phase transition. In this case, the number of species $n$ must 
be about $5\sim 10$ in order to account for $\Delta N_{\rm eff} \sim 1$. 
We need some explanation for a light mass of these particles. 
As noted in the footnote~\ref{ft}, one possibility is a non-Abelian gauge symmetry. Another
one is the supersymmetry. In the SUSY limit, the gravitino becomes massless, and
there might be several gravitinos in a supergravity theory with e.g. $N=8$~\cite{Cheung:2010mc}.
If some of the gravitinos are extremely light, they may reach thermal equilibrium
and decouples before the QCD phase transition~\cite{Pierpaoli:1997im,Ichikawa:2009ir}
and they may account for the extra radiation.

In this paper we have investigated various possibilities to account for the extra radiation
$\Delta N_{\rm eff}\sim 1$ both at the BBN and CMB epochs,
which is  suggested by recent observations. While it seems difficult to significantly improve the constraint on $N_{\rm eff}$ from the helium abundance determination,
the forthcoming Planck result will tell us about $\Delta N_{\rm eff}$ with an accuracy of $\sim 0.3$
at 68\% C.L.~\cite{Ichikawa:2008pz}. Thus the existence of extra radiation will be confirmed or disfavored by the Planck satellite.
Assuming that there is indeed extra radiation and that it is
composed of a light and relativistic unknown particles $X$ which was once in thermal
equilibrium, we have found that a chiral fermion coupled to the SM fermions through an interaction (\ref{4f}) is a viable candidate.
Such a chiral fermion also satisfies the current experimental and astrophysical constraints.

Interestingly, our model of the chiral fermion has implications for the collider experiments. 
Let us comment on possible signatures at the LHC.
In the  model, there exists a new heavy gauge boson $A_H$
of mass of $\mathcal O(1){\rm \,TeV}$
which couples the light chiral fermion and and the SM fermions.
The search strategy for such a gauge boson is same as the $Z'$ boson search.
The Tevatron experiment put the most stringent constraint on the $A_H$ boson mass,
which depends on the coupling $g_{Aff}$~\cite{Nakamura:2010zzi}.
According to the ATLAS study~\cite{Aad:2009wy}, the $A_H$ boson mass of 3\,TeV
is within the reach of 5$\sigma$ discovery with integrated luminosity of $10\,{\rm fb}^{-1}$.

As a concrete realization of such a chiral fermion model, we have constructed a $E_6$-inspired model.
In this model, the above fermion $X$ is identified with a SU(5) singlet fermion $\Psi_{\bf 1}$
in the {\bf 27} representation of $E_6$ and
there also exist  long-lived colored particles $\Psi_{\bf 10}^{\rm color}$ with mass of $\mathcal O(1)$TeV.
The production cross section of such a particle at the LHC is around
$10~{\rm fb}$ for the 1\,TeV colored particle, 
and $10^{-3}~{\rm fb}$ for the 3\,TeV colored particle~\cite{Cheung:2002uz}.
Once produced, it leaves characteristic signatures on detectors depending on 
whether a hadronized particle has an electric charge or not~\cite{Fairbairn:2006gg}, as well as
on whether the particle decays inside the detector or not.
Hence such a long-lived colored particle with mass of a few TeV may be within the reach of LHC.

Thus, if the existence of the extra radiation is confirmed by the Planck satellite,
the LHC may be able to discover signatures of a new gauge boson or a long-lived colored particle
with mass of a few TeV.

%%%%%%%%%%%%%%%%%%%%%%%%%%%%%%%%%%%%
\section*{Acknowledgment}
%%%%%%%%%%%%%%%%%%%%%%%%%%%%%%%%%%%%

FT thanks Javier Redondo and Georg Raffelt for pointing out the relevance of the supernova cooling argument.
TTY thanks K. Izawa for a useful discussion on broken higher N SUSY theories.
FT thanks Alejandro Ibarra, Mathias Garny and the theory group of Technical University Munich
for the warm hospitality while the present work was finalized. 
This work was supported by the Grant-in-Aid for Scientific Research on 
Innovative Areas (No. 21111006) [KN and FT],  Scientific Research (A)
(No. 22244030 [FT] and 22244021 [TTY]), and JSPS Grant-in-Aid for Young Scientists (B) (No. 21740160) [FT]. 
This work was also supported by World Premier
International Center Initiative (WPI Program), MEXT, Japan.

%%%%%%%%%%%%%%%%%%%%%%%%%%%%%%%%%%%%%%%%%%%%%%

\end{document}